\documentclass[10pt,twocolumn,twoside]{IEEEtran}

\usepackage[usenames,dvipsnames]{xcolor}
\usepackage{fancyhdr,amssymb,amsmath, graphicx, listings,float,subfig,enumerate,epstopdf,color,setspace,bm,cite,setspace,amsthm,todonotes, wrapfig, cite}
\def\dist{\operatorname{ dist }}
\def\vol{\operatorname{ vol }}

\newcommand{\aee}{{\cal A}}
\renewcommand{\qed}{\hfill\qedsymbol}

\newcommand{\R}{\mathbb{R}}

\newcommand{\Z}{\mathbb{Z}}
\newcommand{\N}{\mathbb{N}}

\newcommand{\E}{\mathbb{E}}

\renewcommand{\l}{{\ell}}
\renewcommand{\a}{\bar{a}}

\newcommand{\m}[1]{\boldsymbol{ #1 }}
\newcommand{\tb}[1]{\textbf{#1}}
\newcommand{\eps}{\varepsilon}
\newcommand{\given}{\;\bigm\vert\;}
\newcommand{\st}{\,:\,}
\DeclareMathOperator*{\argmax}{arg\,max}

\newcommand{\X}{{\mathcal X}}
\newcommand{\sX}{X}
\newcommand{\x}{{\m{x}}}
\newcommand{\bx}{x}

\renewcommand{\P}\phi
\newcommand{\half}{{1\over 2}}
\newtheorem{theorem}{Theorem}

\newtheorem{lemma}{Lemma}

\theoremstyle{definition}
\newtheorem{defn}{Definition}
\newtheorem{example}{Example}
\newenvironment{noindlist}
 {\begin{list}{\labelitemi}{\leftmargin=0.25em \itemindent=0.25em}}
 {\end{list}}

\begin{document}
\normalsize\title{\LARGE \bf
Fast Convergence in Semi-Anonymous Potential Games\thanks{This research was supported by AFOSR grants \#FA9550-12-1-0359, ONR grant \#N00014-12-1-0643, NSF grant \#ECCS-1351866, the NASA Aeronautics scholarship program, the Philanthropic Educational Organization, and the Zonta International Amelia Earhart fellowship program.}\thanks{We wish to acknowledge conversations with Jinwoo Shin regarding the technical content of this paper, and we thank him for his feedback. }
}

\author{Holly Borowski\thanks{H. Borowski is with the Department of Aerospace Engineering Sciences,  University of Colorado, 429 UCB, Boulder, CO 80309, 719-213-3254. {\texttt{holly.borowski@colorado.edu}.}} and Jason R. Marden\thanks{J. R. Marden is with the Department of Electrical and Computer Engineering, Harold Frank Hall, Rm 5161, University of California, Santa Barbara, 93106, 805-893-2299, {\texttt{jrmarden@ece.ucsb.edu}.  Corresponding author.}}}

\graphicspath{{figures/}}
\maketitle

\setlength{\belowcaptionskip}{-10pt}

\maketitle

\begin{abstract}
Log-linear learning has been extensively studied in both the game theoretic and distributed control literature.  It is appealing for many applications because it often guarantees that the agents' collective behavior will converge in probability to the optimal system configuration. However, the worst case convergence time can be prohibitively long, i.e., exponential in the number of players. Building upon the work in \cite{Shah2010}, we formalize a modified log-linear learning algorithm whose worst case convergence time is roughly linear in the number of players.  We prove this characterization for a class of potential games where agents' utility functions can be expressed as a function of aggregate behavior within a finite collection of populations. Finally, we show that the convergence time remains roughly linear in the number of players even when the players are permitted to enter and exit the game over time.
\end{abstract}

\section{Introduction} 

Game theoretic learning algorithms have gained traction as a design tool for distributed control systems \cite{Marden2008,Zhu2009,Goto2010,Staudigl2012,Fox2010}.  Here, a static game is repeated over time, and agents revise their strategies based on their objective functions and on observations of other agents' behavior.  Emergent collective behavior for such revision strategies has been studied extensively in the literature, e.g., fictitious play \cite{fp1,fp2,jsfp}, regret matching \cite{Hart2000}, and log-linear learning \cite{Alos-Ferrer2010, Blume1993, Shah2010}.  Although many of these learning rules have desirable asymptotic guarantees, their convergence times either remain uncharacterized or are prohibitively long \cite{Ellison2000, Kandori1993,Shah2010,Hart2010}. Characterizing convergence rates is key to determining whether a distributed algorithm is desirable for system control.

In many multi-agent systems, the agent objective functions can be designed to align with the system-level objective function, yielding  a \emph{potential game} \cite{Monderer1996} whose potential function is precisely the system objective function.  Here, the optimal collective behavior of a multi-agent system corresponds to the Nash equilibrium that optimizes the potential function. Hence, learning algorithms which converge to this efficient Nash equilibrium have proven useful for distributed control.

\emph{Log-linear learning} is one algorithm that accomplishes this task \cite{Blume1993}. Log-linear learning is a  {perturbed} best reply process where agents predominantly select the optimal action given their beliefs about other agents' behavior; however, the agents occasionally make mistakes, selecting suboptimal actions with a probability that decays exponentially with respect to the associated payoff loss.  
As noise levels approach zero, the resulting process has a unique stationary distribution with full support on the efficient Nash equilibria.  By designing agents' objective functions appropriately, log-linear learning can be used to define distributed control laws which converge to optimal steady-state behavior in the long run.

Unfortunately, worst-case convergence rates associated with log-linear learning are exponential in the game size \cite{Shah2010}.  This stems from inherent tension between desirable asymptotic behavior and convergence rates.  The tension arises because small noise levels are necessary to ensure that the mass of the stationary distribution lies primarily on the efficient Nash equilibria; however, small noise levels also make it difficult to exit inefficient Nash equilibria,  degrading convergence times.  

Positive convergence rate results for log-linear learning and its variants are beginning to emerge for specific game structures \cite{Montanari2010,Kreindler2011,Shah2010,Arieli2011}.  For example, in \cite{Montanari2010} the authors study the convergence rates of log-linear learning for a class of coordination games played over graphs. They demonstrate that underlying convergence rates are desirable provided that the interaction graph and its subgraphs are sufficiently sparse.  Alternatively, in \cite{Shah2010} the authors introduce a variant of log-linear learning and show that  convergence times grow roughly linearly in the number of players for a special class of congestion games over parallel networks.  They also show that convergence times remain linear in the number of players when players are permitted to exit and enter the game.  Although these results are encouraging, the restriction to parallel networks is severe and hinders the applicability of such results to distributed engineering systems.  

We focus on identifying whether the positive convergence rate results above extend beyond symmetric congestion games over parallel networks to games of a more general structure relevant to distributed engineering systems.  Such guarantees are not automatic because there are many simplifying attributes associated with symmetric congestion games that do not extend in general (see Example~\ref{e:exIntro}).  The main contributions of this paper are as follows:

\vspace{.1cm}
\noindent -- We formally define a subclass of potential games, called \emph{semi-anonymous potential games}, which are parameterized by populations of agents where each agent's objective function can be evaluated using only information regarding the agent's own decision and the aggregate behavior within each population.  Agents within a given population have identical action sets, and their objective functions share the same structural form.  The congestion games studied in \cite{Shah2010} could be viewed as a semi-anonymous potential game with only one population.\footnote{Semi-anonymous potential games can be viewed as a cross between a potential game and a finite population game \cite{Blume1996}.}

\vspace{.1cm}
\noindent -- We introduce a variant of log-learning learning that extends the algorithm in \cite{Shah2010}.  In Theorem~\ref{t:main theorem 1}, we prove that the convergence time of this algorithm grows roughly linearly in the number of agents for a fixed number of populations.  This analysis explicitly highlights the potential impact of system-wide heterogeneity, i.e., agents with different action sets or objective functions, on the convergence rates.  Furthermore, in Example~\ref{e:grow with n} we demonstrate how a given resource allocation problem can be modeled as a semi-anonymous potential game.  

\vspace{.1cm}
\noindent -- We study the convergence times associated with our modified log-linear learning algorithm when the agents continually enter and exit the game.  In Theorem~\ref{t:main theorem 2}, we prove that the convergence time of this algorithm remains roughly linear in the number of agents provided that the agents exit and enter the game at a sufficiently slow rate. 

The forthcoming analysis is similar in structure to the analysis presented in \cite{Shah2010}.  We highlight the explicit differences between the two proof approaches throughout, and directly reference lemmas within \cite{Shah2010} when appropriate.  The central challenge in adapting and extending the proof in \cite{Shah2010} to the setting of semi-anonymous potential games is dealing with the growth of the underlying state space.  Note that the state space in \cite{Shah2010} is characterized by the aggregate behavior of a single population while the state space in our setting is characterized by the Cartesian product of the aggregate behavior associated with several populations.  The challenge arises from the fact that the employed techniques for analyzing the mixing times of this process, i.e., Sobolev constants, rely heavily on the structure of this underlying state space.

\section{Semi-Anonymous Potential Games}

Consider a game with agents $N = \{1,2,\ldots,n\}$. Each agent $i \in N$ has a finite action set denoted by $\aee_i$ and a utility function $U_i : \aee \rightarrow \mathbb{R}$, where $\aee = \prod_{i \in N} \aee_i$ denotes the set of joint actions.   We express an action profile $a \in \aee$ as $(a_i,a_{-i})$ where $a_{-i} = (a_1,\ldots,a_{i-1},a_{i+1},\ldots, a_n)$ denotes the actions of all agents other than agent $i$.  We denote a game $G$ by the tuple $G = \left(N, \{\aee_i\}_{i\in N}, \{U_i\}_{i \in N}\right)$\footnote{For brevity, we refer to $G$ by $G = \left(N, \{\aee_i\}, \{U_i\}\right)$. }.

\begin{defn}\label{d:semi-anon potential}
A game $G$ is a semi-anonymous potential game if there exists a partition ${\cal N} = (N_1, N_2, \dots, N_m)$ of $N$ such that the following conditions are satisfied:

\smallskip

\noindent (i)  For any population $N_{\l} \in {\cal N}$ and agents $i,j \in N_{\l}$ we have $\aee_i = \aee_j$.  Accordingly, we say population $N_{\l}$ has action set $\bar{\aee}_{\l}= \{\a_{\l}^1,\a_{\l}^2,\ldots,\a_{\l}^{s_{\l}}\}$\footnote{We use the notation $\bar{\aee}_\ell$ to represent the action set of the $\ell$th population, whereas $\mathcal{A}_i$ represents the action set of the $i$th agent.} where $s_{\l}$ denotes the number of actions available to population $N_{\l}$.  For simplicity, let $p(i) \in \{1, \dots, m\}$ denote the index of the population associated with agent $i$.  Then, $\aee_i = \bar{\aee}_{p(i)}$ for all agents $i \in N$. 

\smallskip

\noindent(ii) For any population $N_{\l} \in {\cal N}$, let 
\begin{equation}
X_{\l} = \left\{\left({v_{\l}^1\over n},{v_{\l}^2\over n},\ldots,{v_{\l}^{s_{\l}}\over n}\right) \geq \mathbf{0} \st \sum_{k=1}^{s_{\l}} v_{\l}^k = |N_{\l}| \right\}
\end{equation}
represent all possible aggregate action assignments for the agents within population $N_{\l}$.  Here, the utility function of any agent $i \in N_{\l}$ can be expressed as a lower-dimensional function of the form $\bar{U}_i : \bar{\aee}_{p(i)} \times X \rightarrow \mathbb{R}$ where $X = X_{1} \times \dots \times X_m$.  More specifically, the utility associated with agent $i$ for an action profile $a = (a_i,a_{-i}) \in \aee$ is of the form $$U_i(a) = \bar{U}_i(a_i,a|_{X})$$ where
\begin{eqnarray}
a|_{X} &=& (a|_{X_1}, a|_{X_2},\ldots,a|_{X_m})\in X, \\
a|_{X_j} &=& \frac{1}{n}\left\{\left|\{j\in N_{\l} \st a_j = \a_{\l}^k\}\right|\right\}_{k=1,\ldots,s_{\l}}.
\end{eqnarray}
The operator $\cdot|_X$ captures each population's aggregate behavior in an action profile $\cdot$. 

\smallskip

\noindent(iii) There exists a potential function $\phi: X \to\R$ such that for any $a \in \aee$ and agent $i \in N$ with action $a_i' \in \aee_i$,
\small
\begin{equation}U_i(a) - U_i(a_i^{\prime},a_{-i}) = \phi(a|_X) - \phi((a_i^{\prime},a_{-i})|_X).\end{equation}
\normalsize
\end{defn}

\smallskip

\noindent If each agent $i \in N$ is alone in its respective partition, the definition of semi-anonymous potential games is equivalent to that of exact potential games in \cite{Monderer1996}.

\begin{example}[Congestion Games \cite{Beckmann1956}]
Consider a congestion game with players $N = \{1,\ldots,n\}$ and roads $R = \{r_1,r_2,\ldots,r_k\}$. Each road $r\in R$ is associated with a congestion function $C_{r}:\Z_+\to \R$, where $C_r(k)$ is the congestion on road $r$ with $k$ total users.
The action set of each player $i \in N$ represents the set of paths connecting player $i$'s source and destination, and has the form $\mathcal{A}_i\subseteq 2^R$.  The utility function of each player $i \in N$ is given by
$$U_i(a_i,a_{-i}) = -\sum_{r\in a_i} C_r(|a|_r),$$
where $|a|_r = |\{j\in N\st r\in a_j\}|$ is the number of players in joint action $a$ whose path contains road $r$.  This game is a potential game with potential function $\phi:\mathcal{X}\to \R$
\begin{equation}
\phi(a|_X) = -\sum_{r\in R}\sum_{k=1}^{|a|_r}C_r(k).
\end{equation}

When the players' action sets are symmetric, i.e., $\aee_i = \aee_j$ for all agents $i,j \in N$, then a congestion game is a semi-anonymous potential game with a single population.  Such games, also referred to as anonymous potential games, are the focus of \cite{Shah2010}.  When the players' action sets are asymmetric, i.e., $\aee_i \neq \aee_j$ for at least one pair of agents $i,j \in N$, then a congestion game is a semi-anonymous potential game where populations consist of agents with identical path choices.  The results in \cite{Shah2010} are not proven to hold for such settings.  
\end{example}

The following example highlights issues that arise when transitioning from a single population to multiple populations.  

\begin{example}\label{e:exIntro}
Consider a resource allocation game with $n$ players and three resources, $R = \{r_1,r_2,r_3\}.$  Let $n$ be even and divide players evenly into populations $N_1$ and $N_2.$ Suppose that players in $N_1$ may select exactly one resource from $\{r_1,r_2\}$, and players in $N_2$ may select exactly one resource from $\{r_2,r_3\}.$  The welfare garnered at each resource depends on how many players have selected that resource; the resource-specific welfare functions are 
\begin{eqnarray*}
W_{r_1}(k) &=&2k, \\ 
W_{r_2}(k) &=& \min\left\{3k,{3\over 2}n\right\},\\
W_{r_3}(k) &= & k.
\end{eqnarray*}
where $k\in \{0,1,\ldots,n\}$ represents the number of agents selecting a given resource. The total system welfare is $$W(a)=\sum_{r \in R} W_r(|a|_r)$$
for any $a\in \mathcal{A}$, where $|a|_r$ represents the number of agents selecting resource $r$ under action profile $a$.  Assign each agent's utility according to its marginal contribution to the system-level welfare: for agent $i$ and action profile $a$ 
\begin{equation}\label{e:MC util}
U_i(a) = W(a) - W(\emptyset,a_{-i})
\end{equation}
where $\emptyset$ indicates that player $i$ did not select a resource.  The marginal contribution utility in (\ref{e:MC util}) ensures that the resulting game is a potential game with potential function $W$ \cite{Wolpert1999}.  

If the agents had symmetric action sets, i.e., if $\aee_i = \{r_1, r_2, r_3\}$ for all $i \in N$, then this game has exactly one Nash equilibrium with $n/2$ players at resource $r_1$ and $n/2$ players at resource $r_2.$  This Nash equilibrium corresponds to the optimal allocation.  

In contrast, the two population scenario above has many Nash equilibria, two of which are: (i) an optimal Nash equilibrium in which all players from $N_1$ select resource $r_1$ and all players from $N_2$ select resource $r_2,$ and (ii) a suboptimal Nash equilibrium in which all players from $N_1$ select resource $r_2$ and all players from $N_2$ select resource $r_3$.  This large number of equilibria will significantly slow any equilibrium selection process, such as log-linear learning and its variants.

\end{example}

\section{Main Results}

Example~\ref{e:exIntro} invites the question: can a small amount of heterogeneity break down the fast convergence results of \cite{Shah2010}? In this section, we present a variant of log-linear learning \cite{Blume1993} that extends the algorithm for single populations in \cite{Shah2010}.  In Theorem~\ref{t:main theorem 1} we prove that for any semi-anonymous potential game our algorithm ensures (i) the potential associated with asymptotic behavior is close to the maximum and (ii) the convergence time grows roughly linearly in the number of agents for a fixed number of populations.  In Theorem~\ref{t:main theorem 2} we show that these guarantees still hold when agents are permitted to enter and exit the game.  An algorithm which converges quickly to the potential function maximizer is useful for multi-agent systems because agent objective functions can often be designed so that the potential function is identical to the system objective function as in Example~\ref{e:exIntro}.

\subsection{Modified Log-Linear Learning}\label{s:alg description}
The following modification of the log-linear learning algorithm extends the algorithm in \cite{Shah2010}.
Let $a(t)\in\mathcal{A}$ be the joint action at time $t \geq 0$.  Each agent $i \in N$ updates its action upon ticks of a Poisson clock with rate $\alpha n/z_i(t)$, where 
$$z_i(t) = |\{k\in N_{p(i)} \st a_k(t) = a_i(t)\}|,$$
and $\alpha > 0$ is a design parameter which dictates the expected total update rate.  
A player's update rate is higher if he is not using a common action within his population. To continually modify his clock rate, each player must know the value of $z_i(t)$, i.e., the number of players within his population sharing his action choice, for all $t\in \R.$  In many cases, agents also need this information to evaluate their utilities, e.g., when players' utilities are their marginal contribution to the total welfare, as in Example~\ref{e:exIntro}. 
   
When player $i$'s clock ticks, he chooses action $a_i\in \bar{\mathcal{A}}_{p(i)} = \mathcal{A}_i$ probabilistically according to
\begin{align}\label{e:logit response}
{\rm Prob}[a_i(t^+) = a_i\given a(t)]  &=  \frac{e^{\beta U_i(a_i,a_{-i}(t))}}{\sum_{a_i^{\prime} \in\mathcal{A}_i}e^{\beta U_i(a_i^\prime,a_{-i}(t))}} \nonumber\\
&=  \frac{e^{\beta\P(a(t)|_\X)}}{\sum_{a_i^{\prime} \in\mathcal{A}_i} e^{\beta\P((a_i^{\prime},a_{-i}(t))|_\X)}},
\end{align}
for any $a_i\in \mathcal{A}_i,$ where $a_i(t^+)$ indicates the agent's revised action and $\beta$ is a design parameter that determines how likely an agent is to choose a high payoff action. As $\beta\to \infty$, payoff maximizing actions are chosen, and as $\beta\to 0$, agents choose from their action sets with uniform probability.  The new joint action is of the form $a(t^+) = (a_i(t^+), a_{-i}(t))\in \mathcal{A}$, where $t\in\R^+$ is the time immediately before agent $i$'s update occurs. For a discrete time implementation of this algorithm and a comparison with the algorithm in \cite{Shah2010}, please see Appendix~\ref{a:M defn}.

The expected number of updates per second for the continuous time implementation of our modified log-linear learning algorithm is lower bounded by $m\alpha n$ and upper bounded by $(|\bar{\aee}_1| + \dots + |\bar{\aee}_m|) \alpha n$.  
To achieve an expected update rate at least as fast as the standard log-linear learning update rate, i.e., at least $n$ per second, we set $\alpha \geq 1/m$.  These dynamics define an ergodic, reversible Markov process for any $\alpha>0$.  

\subsection{Semi-Anonymous Potential Games}\label{s:main theorem 1}

Theorem~\ref{t:main theorem 1} bounds the convergence time for modified log-linear learning in a semi-anonymous potential game and extends the results of \cite{Shah2010} to semi-anonymous potential games. For notational simplicity, define $s:= |\cup_{j= 1}^m \overline{\mathcal{A}}_j|.$  

\begin{theorem}\label{t:main theorem 1}
Let $G = (N,\{\mathcal{A}_i\},\{U_i\})$ be a semi-anonymous potential game with aggregate state space $X$ and potential function $\P:X\to [0,1].$ Suppose agents play according to the modified log-linear learning algorithm described above, and the following conditions are met:         

\noindent (i)  The potential function is $\lambda$-Lipschitz, i.e., there exists $\lambda \geq 0$ such that
\begin{equation*}
|\P(x) - \P(y)|\leq\lambda\|x-y\|_1,\quad \forall x,y\in  X.
\end{equation*}

\noindent(ii) The number of players within each population is sufficiently large:
$$\sum_{i=1}^m |N_i|^2\geq \sum_{i=1}^m |\bar{\aee}_i| - m.$$  

\noindent For any fixed $\eps\in (0,1)$, if $\beta$ is sufficiently large, i.e., 
\begin{equation}\label{e:beta lb}
\beta\geq\max\left\{{4m(s-1)\over\eps}\log 2ms,{4m(s-1)\over\eps}\log{8ms \lambda\over\eps}\right\},
\end{equation}
then
\begin{equation}\label{e:expected val}
\E[\P(a(t)|_{X})]\geq\max_{x\in X}\P(x)-\eps
\end{equation}
for all
\small
\begin{align}
t&\geq \frac{2^{2ms}c_1 e^{3\beta}m(m(s-1))!^2 n}{4\alpha}\times\nonumber\\
&\quad\quad\quad \Biggl(\log\log (n+1)^{ms - m} +\log\beta+ 2\log{1\over\eps}\Biggr)\label{e:time requirement}
\end{align}
\normalsize
where $c_1$ is a constant that depends only on $s$.  
\end{theorem}

We prove Theorem~\ref{t:main theorem 1} in Appendix~\ref{a:theorem 1 proof}.  This theorem explicitly highlights the role of system heterogeneity, i.e., $m>1$ distinct populations, on convergence times of the process.  For the case when $m=1$, Theorem~\ref{t:main theorem 1} recovers the results of \cite{Shah2010}.  Observe that for a fixed number of populations, the convergence time grows as $n\log\log n$.  Furthermore, note that a small amount of system heterogeneity does not have a catastrophic impact on worst-case convergence times as suggested by Example~\ref{e:exIntro}.

It is important to note that our bound is exponential in the number of populations and in the total number of actions. Therefore our results do not guarantee fast convergence with respect to these parameters. However, our convergence rate bounds may be conservative in this regard. Furthermore, as we will show in Section~\ref{s:examples}, a significantly smaller value of $\beta$ may often be chosen in order to further speed convergence while still retaining the asymptotic properties guaranteed in \eqref{e:expected val}.

\subsection{Time Varying Semi-Anonymous Potential Games}

In this section, we consider a trajectory of semi-anonymous potential games to model the scenario where agents enter and exit the system over time,
$$\mathcal{G} = \{G^t\}_{t\geq 0} =  \{N^t,\{\mathcal{A}_i^t\}_{i\in N^t},\{U_i^t\}_{i\in N^t}\}_{t\geq 0}$$
where, for all $t\in \R^+$, the game $G^t$ is a semi-anonymous potential game, and the set of \emph{active} players, $N^t$, is a finite subset of $\N.$  We refer to each agent $i\in \N\setminus N^t$ as \emph{inactive}; an inactive agent has action set $\mathcal{A}_i^t = \emptyset$ at time $t$. Define $\X :=\cup_{t\in \R^+}X^t$, where $X^t$ is the finite aggregate state space corresponding to game $G^t.$
At time $t$, denote the partitioning of players per Definition~\ref{d:semi-anon potential}  by $\mathcal{N}^t = \{N_1^t,N_2^t,\ldots,N_m^t\}$. We require that there is a fixed number of populations, $m$, for all time, and that the $j$-th population's action set is constant, i.e., $\forall j\in \{1,2,\ldots,m\},\;\forall t_1, t_2 \in \R^+$, $\bar{\mathcal{A}}^{t_1}_j = \bar{\mathcal{A}}^{t_2}_j.$ We write the fixed action set for players in the $j$-th population as  $\bar{\mathcal{A}}_j$.

\begin{theorem}\label{t:main theorem 2}
Let $\mathcal{G}$ be a trajectory of semi-anonymous potential games with state space $\X$ and time-invariant potential function $\P:\X\to [0,1]$.
Suppose agents play according to the modified log-linear learning algorithm and Conditions (i) and (ii) of Theorem~\ref{t:main theorem 1} are satisfied. Fix $\eps\in (0,1)$, assume the parameter $\beta$ satisfies (\ref{e:beta lb}) and the following additional conditions are met:

\noindent(iii) for all $t\in \R^+,$ the number of players satisfies: 
\begin{equation}\label{e:num players}
|N^t| \geq \max\left\{\frac{4\alpha m e^{-3\beta}}{2^{2ms}c_1m^2(m(s-1))!^2},2\beta\lambda+1   \right\},
\end{equation}

\noindent(iv) there exists $k>0$ such that
\begin{equation}\label{e:pop sizes}
|N_i^t| \geq |N^t|\mathop{/} k,\quad \forall i\in \{1,2,\ldots,m\},\;\forall t\in\R^+,
\end{equation}

\smallskip

\noindent (v) there exists a constant
\begin{equation}\label{e:lambda}
\Lambda\geq 8c_0\eps^{-2}e^{3\beta}(6\beta\lambda+e^\beta k(s-1))
\end{equation}
such that for any $t_1,t_2$ with $|t_1 - t_2|\leq \Lambda,$
\begin{equation}
\left|\left\{i\in N^{t_1} \cup N^{t_2} \st \mathcal{A}_i^{t_1}\neq \mathcal{A}_i^{t_1}\right\}\right|\leq 1,
\end{equation}
and, if $ i \in N^{t_1} \cap N^{t_2}$, then $ i \in N_j^t$ for some $j \in \{1, \dots, m\}$ and for all time $t \in [t_1, t_2],$ i.e., agents may not switch populations over this interval.
Here, $c_0$ and $c_1$ do not depend on the number of players, and hence the constant $\Lambda$ does not depend on $n$.
\smallskip

\noindent Then,
\begin{equation}\label{e:expot}
\E[\P(a(t)|_{\X})]\geq \max_{x\in X(t)}\P(x) - \eps
\end{equation}
for all 
\begin{equation}\label{e:t ub}
t\geq |N^0| e^{3\beta}c_0\left({(ms-m)!\log(|N^0|+2)+\beta\over\eps^2}\right).
\end{equation}
\end{theorem}

Theorem~\ref{t:main theorem 2} states that, if player entry and exit rates are sufficiently slow as in Condition~(v), then the convergence time of our algorithm is roughly linear in the number of players. However, the established bound grows quickly with the number of populations. Note that selection of parameter $\beta$ impacts convergence time, as reflected in \eqref{e:t ub}: larger $\beta$ tends to slow convergence. However, the minimum $\beta$ necessary to achieve an expected potential near the maximum, as in \eqref{e:expot}, is independent of the number of players, as given in (\ref{e:beta lb}). The proof of Theorem~\ref{t:main theorem 2} follows a similar structure to the proof of Theorem 4 in \cite{Shah2010} and is hence omitted for brevity. The significant technical differences arise due to differences in the size of the state space when $m>1$. These differences give rise to Condition (iv) in our theorem. 

\section{Illustrative Examples }\label{s:examples}

In this section, we consider resource allocation games with a similar structure to Example~\ref{e:exIntro}. In each case, agents' utility functions are defined by their marginal contribution to the system welfare, $W$, as in \eqref{e:MC util}. Hence, each example is a potential game with potential function $W$.

Modified log-linear learning defines an ergodic, continuous time Markov chain; we denote its transition kernel by $P$ and its stationary distribution by $\pi.$ For relevant preliminaries on Markov chains, please refer to Appendix~\ref{a:MchainPrelims},  and for a precise definition of the transition kernel and stationary distribution associated with modified log-linear learning, please refer to Appendices~\ref{a:M defn} and \ref{a:theorem 1 proof}.

Unless otherwise specified, we consider games with $n$ players distributed evenly into populations $N_1$ and $N_2$. There are three resources, $R = \{r_1,r_2,r_3\}$. Players in population $N_1$ may choose a single resource from $\{r_1,r_2\}$ and players in population $N_2$ may choose a single resource from $\{r_2,r_3\}.$ We represent a state by 
\begin{equation}x = \left(x_1^1,x_2^1,x_2^2,x_3^2\right),\label{e:states}\end{equation}
where $nx_1^1$ and $nx_2^1$ are the numbers of players from $N_1$ choosing resources $r_1$ and $r_2$.  Likewise, $nx_2^2$ and $nx_3^2$ are the numbers of players from $N_2$ choosing resources $r_2$ and $r_3$ respectively.  Welfare functions for each resource depend only on the number of players choosing that resource, and are specified in each example. The system welfare for a given state is the sum of the welfare garnered at each resource, i.e.,
\begin{equation*}
W(x) = W_{r_1}(nx_1^1) + W_{r_2}(n(x_2^1+x_2^2)) + W_{r_3}(nx_3^2).
\end{equation*}
Player utilities are their marginal contribution to the total welfare, $W$, as in \eqref{e:MC util}.

In Example~\ref{e:compare to SS}, we directly the compute convergence times as in Theorem~\ref{t:main theorem 1}: 
\begin{equation}\label{e:example conv time}
\min\{t\st\E_{P^t(y,\cdot)}W(x)\geq \max_{x\in X} W(x) - \eps,\,\forall y\in X\},
\end{equation}
for modified log-linear learning, the variant of \cite{Shah2010}, and standard log-linear learning. This direct analysis is possible due to the example's relatively small state space.

\begin{example}\label{e:compare to SS}

Here, we compare convergence times of our log-linear learning variant, the variant of \cite{Shah2010}, and standard log-linear learning. The transition kernels for each process are described in detail in Appendix~\ref{a:M defn}.

Starting with the setup described above,
we add a third population, $N_3$. Agents in population $N_3$ contribute nothing to the system welfare and may only choose resource $r_2.$ Because the actions of agents in population $N_3$ are fixed, we represent states by aggregate actions of agents in populations $N_1$ and $N_2$ as in \eqref{e:states}.
The three resources have the following welfare functions for each $x = \left(x_1^1,x_2^1,x_2^2,x_3^2\right)\in\sX$:
\begin{align*}
W_{r_1}(x) &= 2nx_1^1,\\
W_{r_2}(x) &= \min\left\{3(nx_1^1+nx_1^2),\frac{3}{2}(nx_2^1+nx_2^2)\right\},\\
W_{r_3}(x) &= nx_3^2.\label{e:ex welfare functions}
\end{align*}
Our goal in this example is to achieve an expected total welfare that is within 98\% of the maximum welfare.

We fix the number of players in populations $N_1$ and $N_2$ at $n_1 = n_2 = 7,$ and vary the number of players in population $n_3$ to examine the sensitivity of each algorithm's convergence rate to the size of $N_3$.

In our variant of log linear learning, increasing the size of population $N_3$ does not change the probability that a player from population $N_1$ or $N_2$ will update next.
However, for standard log-linear learning and for the variant in \cite{Shah2010}, increasing the size of population $N_3$ significantly decreases the probability that players from $N_1$ or $N_2$ who are currently choosing resource $r_2$ will be selected for update.\footnote{Recall that in our log-linear learning variant and the one introduced in \cite{Shah2010}, an updating player chooses a new action according to \eqref{e:logit response}; the algorithms differ only in agents' update rates. In our algorithm, an agent $i$ in population $N_j$'s update rate is $\alpha n\mathop{/}z_i^j(t),$ where $z_i^j(t)$ is the number of agents from population $j$ playing the same action as agent $i$ at time $t.$ In the algorithm in \cite{Shah2010}, agent $i$'s update rate is $\alpha n\mathop{/}\tilde{z}_i(t),$ where $\tilde{z}_i(t)$ is the \emph{total} number of agents playing the same action as agent $i$.}

We select $\beta$ in all cases so that, as $t\to\infty$, the expected welfare associated with the resulting stationary distribution is within 98\% of its maximum. Then we examine the time it takes to come within $\eps = 0.05$ of this expected welfare. We multiply convergence times by the number of players, $n$, to analyze the expected number of updates required to reach the desired welfare. These numbers represent the convergence times when the expected total number of updates per unit time is held constant as $n$ increases. Table~\ref{t:comparison table} depicts $\beta$ values and expected numbers of updates.

For both log-linear learning and our modification, the required $\beta$ to reach an expected welfare within 98\% of the maximum welfare is independent of $n_3$ and can be computed using the expressions 
\begin{align}
\pi_x^{\rm LLL} &\propto e^{\beta W(x)}{n_1 \choose nx_1^1, nx_2^1}{n_2 \choose nx_2^2, nx_3^2},\\
\text{ and }\pi_x^{\rm MLLL} &\propto e^{\beta W(x)}.
\end{align}
These stationary distributions can be verified using reversibility arguments with the standard and modified log-linear learning probability transition kernels, defined in \cite{Shah2010} and Appendix~\ref{a:M defn} respectively. 
Unlike standard log-linear learning and our variant, the required $\beta$ to reach an expected welfare of 98\% of maximum for the log-linear learning variant of \cite{Shah2010} does change with $n_3.$ For each value of $n$, we use the probability transition matrix to  determine the necessary values of $\beta$ which yield an expected welfare of 98\% of its maximum.

Our algorithm converges to the desired expected welfare in fewer updates than both alternate algorithms for all tested values of $n_3$, showing that convergence rates for log linear learning and the variant from \cite{Shah2010} are both more sensitive to the number of players in population 3 than our algorithm.\footnote{A high update rate for players in population $N_3$ was undesirable because they contribute no value. While this example may seem contrived, mild variations would exhibit similar behavior. For example, consider a scenario in which a relatively large population that contributes little to the total welfare may choose from multiple resources.}

\begin{table*}[!]
\begin{center}
\setlength{\tabcolsep}{15pt}
\renewcommand{\arraystretch}{1.25}
\begin{tabular}{c|c|c|c|c}
\tb{Algorithm} & $n_3$ &$\beta$ &	Expected welfare & Expected \# updates \\\hline\hline
\tb{Standard Log Linear Learning}	&1		&3.77 	&98\%		&9430		\\
									&5		&3.77	&98\%			&11947		\\
									&50		&3.77	&98\%		&40250		\\
									&500	&3.77	&98\%		&323277		\\\hline
\tb{Log Linear Learning Variant from \cite{Shah2010}} 
									&1		&2.39	&98\%		&1325		\\
									&5		&2.44	&98\%		&1589		\\
									&50		&2.83	&98\%		&3342		\\
									&500	&3.72	&98\%		&15550		\\\hline
\tb{Our Log Linear Learning Variant}
									&1		&1.28	&98\%		&743		\\
									&5		&1.28	&98\%		&743		\\
									&50		&1.28	&98\%		&743		\\
									&500	&1.28	&98\%		&743		\\
\end{tabular}
\end{center}
\label{t:comparison table}
\caption{This table corresponds to Example~\ref{e:compare to SS}. Here, there are three populations of agents, $N_1,N_2,$ and $N_3,$ and three resources $r_1,r_2,$ and $r_3.$ Agents in population $N_1$ may choose from resources $r_1$ and $r_2,$ and agents in population $N_2$ may choose from resources $r_2$ and $r_3.$ Agents in population $N_3$ may only choose resource $r_2.$ Welfare functions are given in \eqref{e:ex welfare functions}; population $N_3$ contributes nothing to the overall system welfare. Here, we examine the sensitivity of convergence times to the size of $N_3,$ and keep the sizes of populations $N_1$ and $N_2$ fixed at 7.
The third column of this table shows the values of $\beta$ which yield an expected total welfare within 98\% of the maximum. These values of $\beta$ are constant for standard log-linear learning and for our variant, but grow with $n$ for the algorithm in \cite{Shah2010}. 
The final column shows the expected number of updates to achieve the desired near-maximum welfare. This value is constant for our algorithm, but increases with $n$ for the other two. Global update rates are a design parameter dictated by parameter $\alpha$; selecting a global update rate of $n$ per second ($\alpha = 1/m$), convergence times would be a factor of $n$ smaller than the number of updates shown.}
\end{table*}%

\end{example}

We are able to determine convergence times in Example~\ref{e:compare to SS} using each algorithm's probability transition matrix, $P$, because the state space is relatively small. Here, we directly compute the distance of distribution $\mu(t) = \mu(0)P^t$ to the stationary distributions, $\pi^{\rm LLL}$ and $\pi^{\rm MLLL}$ for the selected values of $\beta,$ where $P$ and $\pi$.  
Examples~\ref{e:grow with n} and \ref{e:SensorTarget}, however, have significantly larger state spaces, making similar computations with the probability transition matrix unrealistic. Thus, instead of computing convergence times as in \eqref{e:example conv time} we repeatedly simulate our algorithm from a worst case initial state and approximate convergence times based on average behavior. This method does not directly give the convergence time of Theorem~\ref{t:main theorem 1}, but the average performance over a sufficiently large number of simulations is expected to reflect expected behavior predicted by the probability transition matrix.

\begin{example}\label{e:grow with n}

In this example we consider a scenario similar the previous example, without the third population. That is, agents are evenly divided into two popultions, $N_1$ and $N_2;$ we allow the total number of agents to vary. Agents in $N_1$ may choose either resource $r_1 $ or $r_2$, and agents in $N_2$ may choose either resource $r_2$ or $r_3.$
We consider welfare functions of the following form:
\begin{align}
W_{r_1}(x) &= \frac{e^ {x_1^1} - 1}{e^ {2}},\label{e:welfares for another example 1}\\
W_{r_2}(x) &= \frac{e^{2x_2^1+2x_3^2} - 1}{e^ {2}},\\
W_{r_3}(x) &= \frac{e^{2.5x_4^2} - 1}{e^ {2}}.\label{e:welfares for another example}
\end{align}
for $x = (x_1^1,x_2^1,x_2^2,x_3^2)\in X.$
Here, the global welfare optimizing allocation is $a_i = r_2$ for all $i\in N$, i.e., $x^{\rm opt} = (0,1/2,1/2,0).$   Similar to Example~\ref{e:exIntro}, this example has many Nash equilibria, two of which are $x^{\rm opt}$ and $x^{\rm ne} = (1/2,0,0,1/2).$

We simulated our algorithm with $\alpha = 1\mathop{/}4$ starting from the inefficient Nash equilibrium, $x^{\rm ne}$.  Here, $\beta$ is chosen to yield an expected steady state welfare equal to 90\% of the maximum. We examine the time it takes the average welfare to come within $\eps = 0.05$ of this expected welfare.

Simulation results are shown in Figure \ref{f:convTimes} averaged over 2000 simulations with $n$ ranging from 4 to 100.  Average convergence times are bounded below by $2n\log\log n$ for all values of $n$, and are bounded above by $4n \log\log n$ when $n>30$.
These results support  Theorem~\ref{t:main theorem 1}.

\begin{figure}[!]
  \centering
    \includegraphics[trim = 0mm 20mm 0mm 20mm, clip, width=0.40\textwidth]{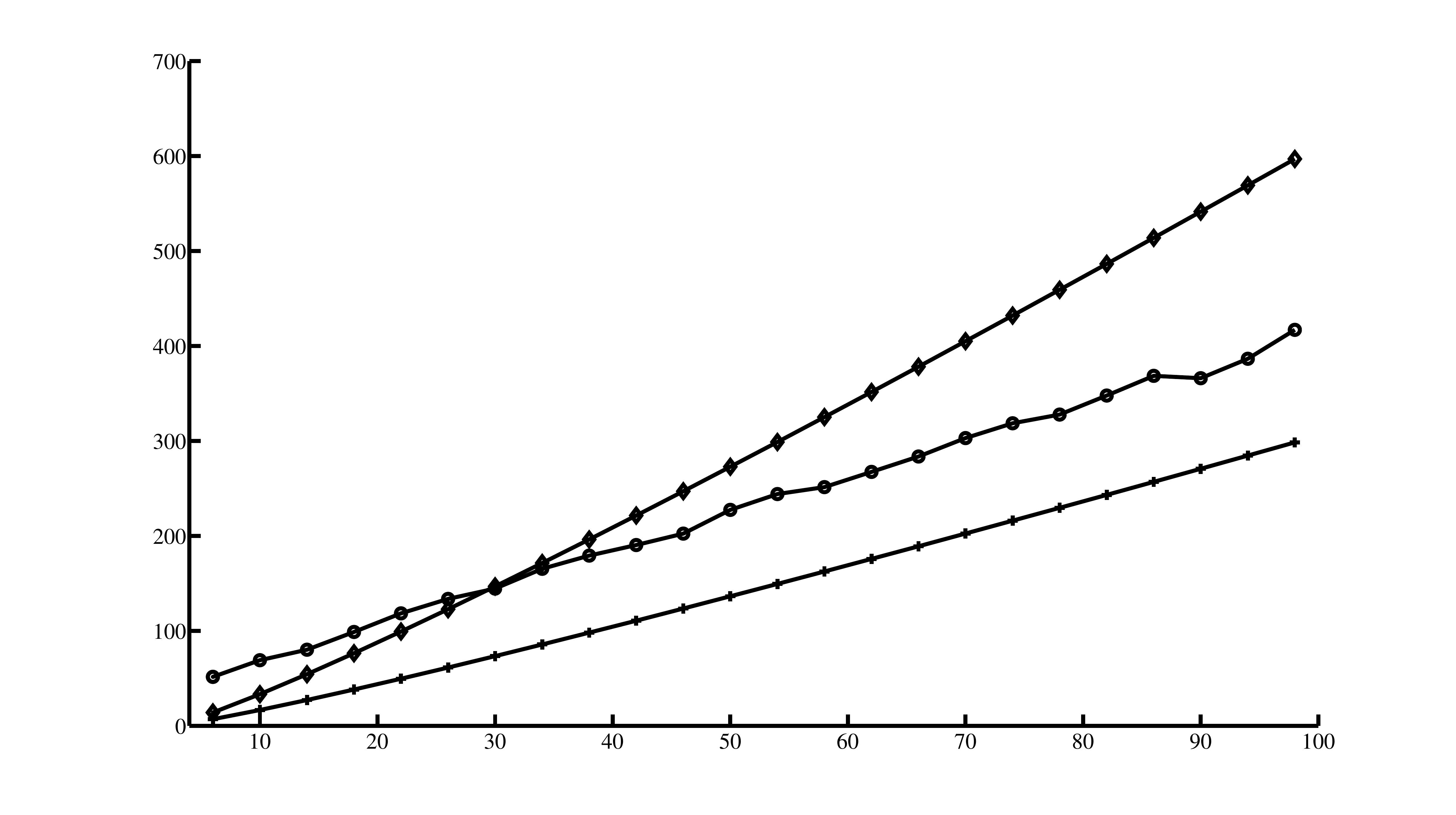}
  \caption{Example~\ref{e:grow with n}, number of players vs. average convergence times. Here, there are two equal-sized populations of agents, $N_1$ and $N_2,$ and three resources $r_1,$ $r_2,$ and $r_3.$ 
Agents in population $N_1$ may choose from resources $r_1$ and $r_2,$ and agents in population $N_2$ may choose from resources $r_2$ and $r_3.$ Welfare functions are given in \eqref{e:welfares for another example}. \label{f:convTimes}}
\end{figure}

\end{example}

\begin{example}\label{s:large action sets}
In this example we investigate convergence times for modified log-linear learning when agents have larger action sets.   We consider the situation where $n$ agents are divided into two populations, $N_1$ and $N_2$. Agents in $N_1$ may choose from resources in $A_1 =\{r_1,r_2,\ldots,r_k\}$, and agents in population $N_2$ may choose from resources in $A_2 = \{r_k,r_{k+1}, \ldots, r_{2k-1}\}.$ That is, each agent may choose from $k$ different resources, and the two populations share resource $r_k$. Suppose resource welfare functions are
\begin{equation}\label{e:large action welfares}W_{r_j}(x) = 
\begin{cases}
x\mathop{/}4n  & \text{if } j\neq k\\
x^2\mathop{/}n^2 &\text{if } j=k,
\end{cases}
\end{equation}
and suppose agents' utilities are given by their marginal contribution to the total welfare, as in \eqref{e:MC util}. We allow $k$ to vary between 5 and 15, and $n$ to vary between 4 and 50. 

The welfare maximizing configuration is for all agents to choose resource $r_k$; however, when all agents in populations $N_1$ and $N_2$ choose resources $r_j$ and $r_{\ell}$ respectively, with $j,\ell\neq k,$ this represents an inefficient Nash equilibrium. Along any path from this type of inefficient Nash equilibrium to the optimal configuration, when $n\geq 4,$ at least $\lceil (n+4)/8\rceil$ agents must make a utility-decreasing decision to move to resource $r_k$. Moreover, the additional resources are all alternative suboptimal choices each agent could make when revising its action; these alternate choices further slow convergence times. Figure~\ref{f:LargeActionSets} shows the average time it takes to reach a configuration whose welfare is 90\% of the maximum, starting from an inefficient Nash equilibrium where all agents in $N_1$ choose resource $r_1$ and all agents in $N_2$ choose resource $r_{2k-1}.$ Parameter $\beta$ is selected so that the expected welfare is at least 90\% of the maximum in the limit as $t\to \infty.$ For each value of $k$, convergence times remain approximately linear in the number of agents, supporting Theorem~\ref{t:main theorem 1}.\footnote{In this example, convergence times appear super-linear in the size of populations' action sets. Note that the bound in \eqref{e:time requirement} is exponential in the the sum of the sizes of each population's action set.  Fast convergence with respect to parameter $s$ warrants future investigation; in particular, convergence rates for our log-linear learning variant may be significantly faster than suggested in \eqref{e:time requirement} under certain mild restrictions on resource welfare functions (e.g., submodularity) or for alternate log-linear learning variants (e.g., binary log-linear learning \cite{Arslan2007, Marden2007a}).}

\begin{figure}[!]
  \centering
    \includegraphics[trim = 40mm 10mm 40mm 20mm, clip, width=0.45\textwidth]{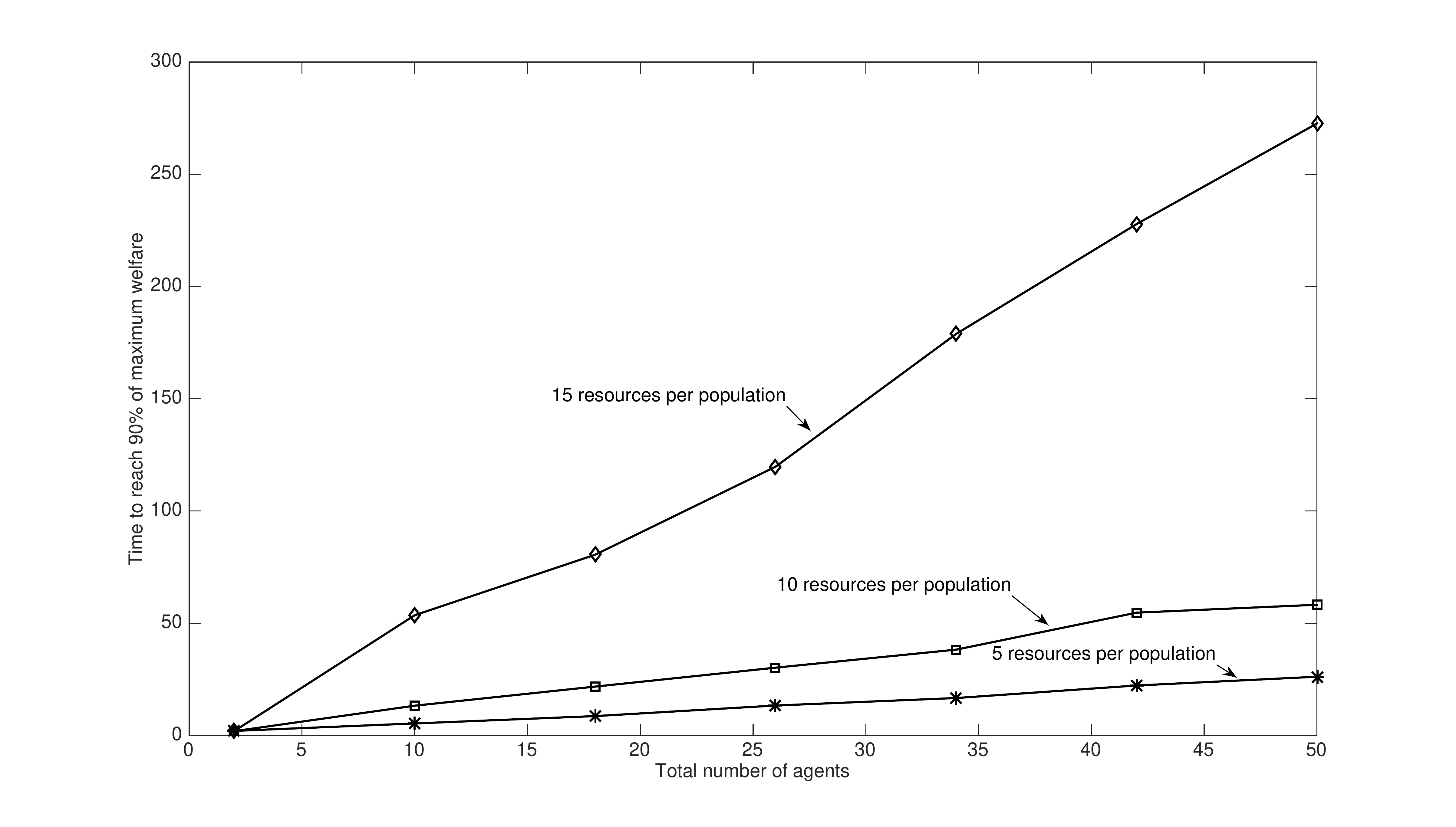}
  \caption{ ~\ref{s:large action sets}, number of agents vs. average time to reach 90\% of the maximum welfare. Agents are separated into two populations, $N_1$ and $N_2.$ Agents in $N_1$ choose from resources $r_1, r_2,\ldots,r_k$, and agents in $N_2$ choose from resources $r_k, r_{k+1}\ldots,r_{2k-1},$ where $k$ varies from 5 to 15. Resource welfare functions are given by \eqref{e:large action welfares}, agent utility functions are given by \eqref{e:MC util}, and average convergence times are taken over 200 simulations. \label{f:LargeActionSets}}
\end{figure}

\end{example}

In Example~\ref{e:SensorTarget} we compare convergence times for standard and modified log-linear learning in a sensor-target assignment problem.

\begin{example}[Sensor-Target Assignment]\label{e:SensorTarget}
In this example, we assign a collection of mobile sensors to four regions. Each region contains a single target, and the sensor assignment should maximize the probability of detecting the targets, weighted by their values. The targets in regions $R = \{r_1,r_2,r_3,r_4\}$ have values
\begin{equation}
v_1 = 1,\quad v_2 = 2,\quad v_3 = 3,\quad v_4 = 4\label{e:area vals}
\end{equation}
respectively. Three  types of sensors will be used to detect the targets: strong, moderate, and weak. Detection probabilities of these three sensor types are:
\begin{equation}p_s = 0.9,\quad p_m = 0.5,\quad p_w = 0.05.\label{e:detection probs}
\end{equation}
The numbers of strong and weak sensors are $n_s = 1$ and $n_m = 5.$  We vary the number of weak sensors, $n_w$.

The expected welfare for area $r_i$ is the detection probability of the collection of sensors located at $r_i$ weighted by the value of target $i$:
\begin{equation*}
W_{r_i}(k_s,k_m,k_w) = v_i\left( 1 - (1 - p_s)^{k_s}(1 - p_m)^{k_m}(1 - p_w)^{k_w}\right),
\end{equation*}
where $k_s,$ $k_m$ and $k_w$ represent the number of strong, moderate, and weak sensors located at region $r_i.$
The total expected welfare for configuration $a$ is
\begin{equation*}
W(a) = \sum_{r\in R} W_r(|a|_r^s,|a|_r^m,|a|_r^w),
\end{equation*}
where $|a|_r^s,|a|_r^m,$ and $|a|_r^w$ are the numbers of strong, moderate, and weak sensors choosing region $r$ in $a$.

We assign agents' utilities according to their marginal contributions to the total welfare, $W$, as in \eqref{e:MC util}. 
Our goal is to reach 98\% of the maximum welfare. We set the initial state to be a worst-case Nash equilibrium.\footnote{The initial configuration is chosen by assigning weak agents to the highest value targets and then assigning strong agents to lower value targets. In particular, agents are assigned in order of weakest to strongest according to their largest possible marginal contribution. This constitutes an inefficient Nash equilibrium. As a similar example, consider a situation with two sensors with detection probabilities $p_1 = 0.5$ and $p_2 = 1$, and two targets with values $v_1 = 2$ and $v_2 = 1.$ The assignment (sensor 1$\to$ target 1, sensor 2$\to$ target 2) is an inefficient Nash equilibrium, whereas the opposite assignment is optimal.  The large state space makes it infeasible to directly compute a stationary distribution, and hence also infeasible to compute values of $\beta$ that will yield precisely the desired expected welfare. Thus, we use simulations to estimate the $\beta$ which yields an expected welfare of 98\% of the maximum.}
 
To approximate convergence times, we simulate each algorithm with the chosen $\beta$ value\footnote{To approximate the value of $\beta$ which yields the desired steady-state welfare of 98\% of maximum, we simulated the standard and modified versions of log-linear learning for $1\times 10^6$ iterations for a range of $\beta$ values. We then selected the $\beta$ which yields an average welfare closest to the desired welfare during the final $5000$ iterations. Note that we could instead set $\beta$ according to \eqref{e:beta lb} for the modified log-linear learning algorithm; however, in order to compare convergence times of modified and standard log-linear learning, we chose $\beta$ to achieve approximately the same expected welfare for both algorithms.} and compute a running average of the total welfare over 1000 simulations. In Figure~\ref{f:SensorTarget} we show the average number of iterations necessary to reach 98\% of the maximum welfare.
 
For small values of $n_w$, standard log-linear learning converges more quickly than our modification, but modified log-linear learning converges faster than the standard version as $n_w$ increases. The difference in convergence times is significant ($\approx 1000$ iterations) for intermediate values of $n_w.$ As the total number of weak sensors increases, (1) the probabilities of transitions along the paths to the efficient Nash equilibrium begin to increase for both algorithms, and (2) more sensor configurations are close to the maximum welfare. Hence, convergence times for both algorithms decrease as $n_w$ increases.

This sensor-target assignment problem does not display worst-case convergence times with respect to the number of agents for either algorithm. However, it demonstrates a situation where our modification can have an advantage over standard log-linear learning. In log-linear learning, the probability that the strong sensor will update next decreases significantly as the number of agents grows. In modified log-linear learning this probability remains fixed. This property is desirable for this particular sensor-target assignment problem, since the single strong sensor contributes significantly to the total system welfare.

\begin{figure}[!]
  \centering
    \includegraphics[trim = 0mm 40mm 0mm 40mm, clip, width=0.40\textwidth]{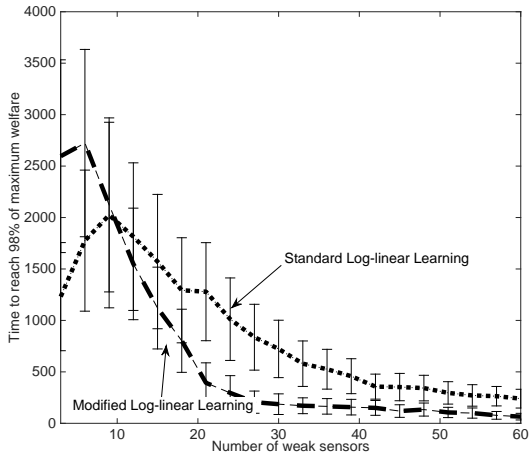}
  \caption{ Example~\ref{e:SensorTarget}, number of weak sensors vs. average convergence times. Here, there are three types of sensors which may choose from four resources. Sensor detection probabilities and resource values are given in \eqref{e:detection probs} and \eqref{e:area vals}. We fix the number of strong and moderate sensors and vary the number of weak sensors. This figure shows the average time it takes for the average welfare to reach 98\% of maximum. The average is taken over 1000 iterations, and convergence times correspond to a global update rate of 1 per second. Error bars show standard deviations of the convergence times.\label{f:SensorTarget}}
\end{figure}

\end{example}

\section{Conclusion}

We have extended the results of \cite{Shah2010} to define dynamics for a class of semi-anonymous potential games whose player utility functions may be written as functions of aggregate behavior within each population.  For games with a fixed number of actions and a fixed number of populations, the time it takes to come arbitrarily close to a potential function maximizer is linear in the number of players. This convergence time remains linear in the initial number of players even when players are permitted to enter and exit the game, provided they do so at a sufficiently slow rate.

\bibliographystyle{plain}
\bibliography{PaperReferences-Journal2014_ConvergenceRates}

\begin{thebibliography}{10}

\bibitem{Alos-Ferrer2010}
C.~Al\'{o}s-Ferrer and N.~Netzer.
\newblock {The logit-response dynamics}.
\newblock {\em Games and Economic Behavior}, 68(2):413--427, 2010.

\bibitem{Arieli2011}
I.~Arieli and H.P. Young.
\newblock {Fast convergence in population games}.
\newblock 2011.

\bibitem{Arslan2007}
G.~Arslan, J.~R. Marden, and J.~S. Shamma.
\newblock {Autonomous vehicle-target assignment: a game theoretical
  formulation}.
\newblock {\em ASME Journal of Dynamic Systems, Measurement and Control},
  129(5):584--596, 2007.

\bibitem{Beckmann1956}
M.~Beckmann, C.B. McGuire, and C.~B. Winsten.
\newblock {\em {Studies in the Economics of Transportation}}.
\newblock Yale University Press, New Haven, 1956.

\bibitem{Blume1993}
L.~Blume.
\newblock {The statistical mechanics of strategic interaction}.
\newblock {\em Games and Economic Behavior}, 1993.

\bibitem{Blume1996}
L.~E. Blume.
\newblock {Population games}.
\newblock {\em The Economy as a Complex Evolving System II}, pages 425--460,
  1996.

\bibitem{Ellison2000}
G.~Ellison.
\newblock {Basins of attraction, long-run stochastic stability, and the speed
  of step-by-step evolution}.
\newblock {\em The Review of Economic Studies}, pages 17--45, 2000.

\bibitem{fp2}
D.~P. Foster and H.~P. Young.
\newblock {On the Nonconvergence of Fictitious Play in Coordination Games}.
\newblock {\em Games and Economic Behavior}, 25(1):79--96, October 1998.

\bibitem{Fox2010}
M.~Fox and J.~Shamma.
\newblock {Communication, convergence, and stochastic stability in
  self-assembly}.
\newblock In {\em 49th IEEE Conference on Decision and Control (CDC)}, December
  2010.

\bibitem{Frieze1998}
A.~Frieze and R.~Kannan.
\newblock {Log-Sobolev inequalities and sampling from log-concave
  distributions}.
\newblock {\em The Annals of Applied Probability}, 1998.

\bibitem{Goto2010}
T.~Goto, T.~Hatanaka, and M.~Fujita.
\newblock {Potential game theoretic attitude coordination on the circle:
  Synchronization and balanced circular formation}.
\newblock {\em 2010 IEEE International Symposium on Intelligent Control}, pages
  2314--2319, September 2010.

\bibitem{Hart2010}
S.~Hart and Y.~Mansour.
\newblock {How long to equilibrium? The communication complexity of uncoupled
  equilibrium procedures}.
\newblock {\em Games and Economic Behavior}, 69:107--126, May 2010.

\bibitem{Hart2000}
S.~Hart and A.~Mas‐Colell.
\newblock {A Simple Adaptive Procedure Leading to Correlated Equilibrium}.
\newblock {\em Econometrica}, 68(5):1127--1150, 2000.

\bibitem{Kandori1993}
M.~Kandori, G.~Mailath, and R.~Rob.
\newblock {Learning, Mutation, and Long Run Equilibria in Games}.
\newblock {\em Econometrica}, 61(1):29--56, 1993.

\bibitem{Kreindler2011}
G.~Kreindler and H.~Young.
\newblock {Fast convergence in evolutionary equilibrium selection}.
\newblock 2011.

\bibitem{jsfp}
J.~Marden.
\newblock {Joint strategy fictitious play with inertia for potential games}.
\newblock {\em IEEE Transactions on Automatic Control}, 54(2):208--220, 2009.

\bibitem{Marden2007a}
J.~Marden, G.~Arslan, and J.~Shamma.
\newblock {Connections between cooperative control and potential games
  illustrated on the consensus problem}.
\newblock {\em Proceedings of 2007 the European Control Conference}, 2007.

\bibitem{Marden2008}
J.~Marden and A.~Wierman.
\newblock {Distributed welfare games}.
\newblock {\em Operations Research}, 61(1):155--168, 2013.

\bibitem{fp1}
D.~Monderer and L.~Shapley.
\newblock {Fictitious Play Property for Games with Identical Interests}.
\newblock {\em Journal of Economic Theory}, (68):258--265, 1996.

\bibitem{Monderer1996}
D.~Monderer and L.~Shapley.
\newblock {Potential games}.
\newblock {\em Games and Economic Behavior}, 14:124--143, 1996.

\bibitem{Montanari2010}
A.~Montanari and A.~Saberi.
\newblock {The spread of innovations in social networks}.
\newblock {\em Proceedings of the National Academy of Sciences}, pages
  20196--20201, 2010.

\bibitem{Montenegro2006}
R.~Montenegro and P.~Tetali.
\newblock {Mathematical Aspects of Mixing Times in Markov Chains}.
\newblock {\em Foundations and Trends in Theoretical Computer Science},
  1(3):237--354, 2006.

\bibitem{Shah2010}
D.~Shah and J.~Shin.
\newblock {Dynamics in Congestion Games}.
\newblock In {\em ACM SIGMETRICS International Conference on Measurement and
  Modeling of Computer Systems}, 2010.

\bibitem{Staudigl2012}
M.~Staudigl.
\newblock {Stochastic stability in asymmetric binary choice coordination
  games}.
\newblock {\em Games and Economic Behavior}, 75(1):372--401, May 2012.

\bibitem{Wolpert1999}
D.~Wolpert and K.~Tumer.
\newblock {An Introduction To Collective Intelligence}.
\newblock Technical report, 1999.

\bibitem{Zhu2009}
M.~Zhu and S.~Martinez.
\newblock {Distributed coverage games for mobile visual sensors (I): Reaching
  the set of Nash equilibria}.
\newblock In {\em Joint 48th IEEE Conference on Decision and Control and 28th
  Chinese Control Conference}, 2009.

\end{thebibliography}

\appendix

\subsection{Markov chain preliminaries}\label{a:MchainPrelims}

A continuous time Markov chain, $\{Z_t\}_{t\geq 0}$, over a finite state space $\Omega$ may be written in terms of a corresponding discrete time chain with transition matrix $M$ \cite{Montenegro2006}, where the distribution $\mu(t)$ over $\Omega$ evolves as:\begin{equation}\label{e:contin time chain}
\mu(t) = \mu(0)e^{t(M-I)} = \mu(0)e^{-t}\sum_{k=0}^\infty {t^kM^k\over k!},\quad t\geq 0
\end{equation}
\normalsize
where we refer to $M$ as the kernel of the process $Z_t$ and $\mu(0) \in \Delta(\Omega)$ is the initial distribution.  
The following definitions and theorems are taken from \cite{Shah2010, Montenegro2006}.  Let $\mu,\nu$ be measures on the finite state space $\Omega.$  Total variation distance is defined as 
\begin{equation}\|\mu - \nu\|_{TV} := \frac{1}{2}\sum_{x\in \Omega}|\mu_x - \nu_x|.\end{equation}
and 
\begin{equation}D(\mu:\nu) := \sum_{x\in\Omega}\mu_x\log\frac{\mu_x}{\nu_x}\end{equation}
is defined to be the relative entropy between $\mu$ and $\nu$. The total variation distance between two distributions can be bounded using the relative entropy: 
\begin{equation}\label{measure ineq}\|\mu - \nu\|_{TV}\leq \sqrt{\frac{D(\mu:\nu)}{2}}\end{equation}
For a continuous time Markov chain with kernel $M$ and stationary distribution $\pi$, the distribution $\mu(t)$ obeys
\begin{equation}\label{e:entropy decay}
D(\mu(t):\pi)\leq e^{-4t\rho(M)}D(\mu(0):\pi),\quad t\geq 0
\end{equation}
where $\rho(M)$ is the Sobolev constant of $M$, defined by 
\begin{equation}\label{e:Sobolev const}
\rho(M) := \inf_{\substack{f: \Omega\to \R \st \\ \mathcal{L}(f)\neq 0}}\frac{\mathcal{E}(f,f)}{\mathcal{L}(f)}
\end{equation}
with 
\begin{align}
\mathcal{E}(f,f) &:= \half \sum_{x,y\in\Omega} (f(x) - f(y))^2 M(x,y)\pi_x\label{e:Eff}\displaybreak[3]\\
\mathcal{L}(f) &:= \E_{\pi}\log {f^2\over \E_{\pi}f^2}.\label{e:Lf}
\end{align}
Here $\E_{\pi}$ denotes the expectation with respect to stationary distribution $\pi$.  For a Markov chain with initial distribution $\mu(0)$ and stationary distribution $\pi$,   the total variation and relative entropy mixing times are 
\begin{align}
T_{TV}(\eps) &:= \min_t\{\|\mu(t) - \pi\|\leq \eps\}\label{e:tv mix time}\\
T_{D}(\eps) &:=\min_t\{D(\mu(t):\pi)\leq\eps\}
\end{align}  
respectively. From \cite{Montenegro2006}, Corollary 2.6 and Remark 2.11, 
\begin{equation*}T_D(\eps)\leq \frac{1}{4\rho(M)}\left(\log\log\frac{1}{\pi_{\min}}+\log\frac{1}{\eps}\right),\end{equation*}
where $\pi_{\min}:= \min_{x\in\Omega} \pi_x.$  Applying \eqref{measure ineq}, 
\small
\begin{align}
T_{TV}(\varepsilon)&\leq T_D(2\eps^2)\displaybreak[3]\nonumber\\
&\leq \frac{1}{4\rho(M)}\left(\log\log\frac{1}{\pi_{\min}}+2\log\frac{1}{\eps}\right).\label{e:mix time bounds}
\end{align}
\normalsize
Hence, a lower bound on the Sobolev constant 
yields an upper bound on the mixing time for the Markov chain.

\subsection{Notation and Problem Formulation: Stationary Semi-Anonymous Potential Games}\label{a:M defn}
The following Markov chain, $M$, over state space $\sX$ is the kernel of the continuous time modified log-linear learning process for stationary semi anonymous potential games.  Define $n_j := |N_j|$ to be the size of population $j$, define $s_j:= |\bar{\mathcal{A}}_j|$, and let $\sigma := \sum_{j = 1}^m s_j.$ Let $e_j^k \in \R^{s_j}$ be the $k$th standard basis vector of length $s_j$ for $k\in \{1,\ldots,s_j\}$.      Finally, let $$x = (\bx_j,\x_-j) = (\bx_1,\bx_2,\ldots,\bx_m)\in \sX,$$
where $\bx_j = (\bx_j^1,\bx_j^2,\ldots,\bx_j^{s_j})$ represents the proportion of players choosing each action within population $j$'s action set.  The state transitions according to:
\begin{noindlist}
\item Choose a population $N_j\in \{N_1,N_2,\ldots,N_m\}$ with probability $s_j/\sigma.$
\item Choose an action $\a_j^k\in \{\a_j^1,\a_j^2,\ldots,\a_j^{s_j}\} =  \bar{\mathcal{A}}_{j}$ with probability $1/s_j.$ 
\item If $\bx_j^k>0$, i.e., at least one player from population $j$ is playing action $\bar{a}_j^k,$ choose $p\in \{p^\prime\in N_j \st \a_{p^\prime} = \a_{j}^k\}$ uniformly at random to update according to \eqref{e:logit response}. That is, transition to $ \left(\bx_j + \frac{1}{n}(e_j^{\ell} - e_j^k),\bx_{-j}\right)$ with probability 
\begin{equation*}\label{e:logit response 2}
\frac{e^{\beta \P\left(\bx_{j} + \frac{1}{n}(e_j^{\ell} - e_j^k),\bx_{-j}\right)}}{\sum_{t =1}^{s_j} e^{\beta\P(\bx_{j} + \frac{1}{n}(e^{\ell}_j - e_j^k), \bx_{-j})}}
\end{equation*}
for each $\ell\in \{1,2,\ldots,s_j\}.$
\footnote{Agents' update rates are the only difference between our algorithm, standard log-linear learning, and the log-linear learning variant of \cite{Shah2010}. In standard log-linear learning, players have uniform, constant clock rates. In our variant and the variant of \cite{Shah2010}, agents' update rates vary with the state. For the algorithm in \cite{Shah2010}, agent $i$'s update rate is $\alpha n\mathop{/}\tilde{z}_i(t),$ where $\tilde{z}_i(t)$ is the \emph{total} number of players selecting the same action as agent $i$. The discrete time kernel of this process is as follows \cite{Shah2010}:
(1) Select an action $a_i\in \cup_{i\in N} A_i$ uniformly at random.
(2) Select a player who is currently playing action $a_i$ uniformly at random. This player updates its action according to \eqref{e:logit response}.
The two algorithms differ when at least two populations have overlapping action sets. }
\end{noindlist}

This defines transition probabilities in $M$ for transitions from state $x = (\bx_{j},\bx_{-j})\in \sX$ to a state of the form $y = \left(\bx_{j} + \frac{1}{n}(e^{\ell}_j - e_j^k),\bx_{-j}\right)\in \sX$ in which a player from population $N_j$ updates his action, so that
\begin{align}
M(x,y) =\frac{e^{\beta \P\left(\bx_{j} + \frac{1}{n}(e^{\ell}_j - e_j^k),\bx_{-j}\right)}}{\sigma\sum_{t =1}^{s_j} e^{\beta\P(\bx_{j} + \frac{1}{n}(e^{t}_j - e_j^k), \bx_{-j})}}\label{e:Markov transition probs}
\end{align}
\normalsize
For a transition of any other form, 
$M(x,y) = 0.$
Applying \eqref{e:contin time chain} to the chain with kernel $M$ and global clock rate $\alpha\sigma n$, modified log-linear learning evolves as
\begin{equation}
\mu(t) = \mu(0)e^{\alpha\sigma n t (M-I)}.
\end{equation}

\noindent\emph{Notation summary for stationary semi-anonymous potential games:}
Let $G = \{N,\{A_i\},\{U_i\}\}$ be a stationary semi-anonymous potential game. The following summarizes the notation corresponding to game $G.$
\begin{itemize}
\item $\sX$ - aggregate state space corresponding to the game $G$
\item $\phi: \sX\to \R$ - the potential function corresponding to game $G$
\item $M$ - probability transition kernel for the modified log-linear learning process
\item $\alpha$ - design parameter for modified log-linear learning which may be used to adjust the global update rate 
\item $\mu(t) = \mu(0)e^{\alpha nt(M-I)}$ - distribution over state space $\sX$ at time $t$ when beginning with distribution $\mu(0)$ and following the modified log-linear learning process
\item $N_j$ - the $j$th population
\item $n_j:= |N_j|$ - the size of the $j$th population
\item $\bar{\mathcal{A}}_j$ - action set for agents belonging to population $N_j$
\item $\a_j^k$ - the $k$th action in population $N_j$'s action set
\item $s:= |\cup_{j= 1}^m \overline{\mathcal{A}}_j|$  - size of the union of all populations' action sets
\item $s_j: = |\bar{\mathcal{A}}_j|$ - size of population $N_j$'s action set
\item $e_j^k\in \R^{s_j}$ - $k$th standard basis vector of length $s_j$
\item $\sigma:= \sum_{j = 1}^m s_j$ - sum of sizes of each population's action set
\item $\pi$ - stationary distribution corresponding to the modified log-linear learning process for game $G$.
\item $(\bx_{j},\bx_{-j}) = (\bx_{1},\bx_{2},\ldots,\bx_{m})\in X$, a state in the aggregate state space, where $\bx_{j} = (\bx_{j}^1,\bx_{j}^2,\ldots,\bx_{j}^{s_j}).$
\end{itemize}

\subsection{Proof of Theorem~\ref{t:main theorem 1}}\label{a:theorem 1 proof}

We require two supporting lemmas to prove Theorem \ref{t:main theorem 1}.  The first  
establishes the stationary distribution for modified log-linear learning as a function of $\beta$ and characterizes how large $\beta$ must be so the expected value of the potential function is within $\eps/2$ of maximum.  The second upper bounds the mixing time to within $\eps/2$ of the stationary distribution for the modified log-linear learning process.

\begin{lemma}\label{l:stationary distribution}\label{l:beta bound}
For the stationary semi-anonymous potential game $G = (N,\mathcal{A}_i,U_i)$  
with state space $\sX$ and potential function $\P: \sX \to [0,1],$
the stationary distribution for modified log-linear learning is
\begin{equation}\label{e:stationary distribution}
\pi_{x}\propto e^{\beta\P(x)}, \quad x\in  \sX
\end{equation}
Moreover, if condition (i) of Theorem~\ref{t:main theorem 1} is satisfied and $\beta$ is sufficiently large as in \eqref{e:beta lb}, then \begin{equation}\label{e:expPot}
\E_\pi[\P(x)]\geq\max_{x\in \sX}\P(x)-\eps/2.
\end{equation}
\end{lemma}

\noindent\emph{Proof:}
The form of the stationary distribution follows from standard reversibility arguments, using \eqref{e:Markov transition probs} and \eqref{e:stationary distribution}.

For the second part of the proof,  define the following:
\small
\begin{align*}
C_\beta & := \sum_{x\in \sX}e^{\beta\P(x)},\displaybreak[3]\\
x^\star &:= \argmax_{x\in \sX}\P(x) \displaybreak[3]\\
 B(x^\star,\delta) &:= \{x\in \sX\st \|x - x^\star\|_1\leq\delta\}
\end{align*}
\normalsize
where $\delta\in[0,1]$ is a constant which we will specify later.  Because $\pi$ is of exponential form with normalization factor $C_\beta$, the derivative of $\log C_\beta$ with respect to $\beta$ is $\E_\pi[\P(x)]$.  Moreover, it follows from \eqref{e:stationary distribution} that $\E_\pi[\P(x)]$ is monotonically increasing in $\beta$, so we may proceed as follows:
\begin{align*}
\E_\pi[\P(x)]	&\geq {1\over\beta}(\log C_\beta - \log C_0)\displaybreak[3]\\
		&=\P(x^\star) +{1\over\beta}\log{\sum_{x\in \sX}e^{\beta(\P(x) - \P(x^\star))}\over| \sX|}\displaybreak[3]\\
		&\stackrel{(a)}{\geq}\P(x^\star)+{1\over\beta}\log{\sum_{x\in B(x^\star,\delta)}e^{-\beta\delta\lambda}\over | \sX|}\displaybreak[3]\\
		&=\P(x^\star)+{1\over\beta}\log{|B(x^\star,\delta)|e^{-\beta\delta\lambda}\over | \sX|}\displaybreak[3]\\
		&=\P(x^\star)-\delta\lambda+{1\over\beta}\log\left({|B(x^\star,\delta)|\over | \sX|}\right)\displaybreak[3]
\end{align*}
where (a) is from the fact that $\P$ is $\lambda$-Lipschitz and the definition of $B(x^\star,\delta)$.  Using intermediate results in the proof of Lemma 6 of \cite{Shah2010}, $|B(x^\star,\delta)|$
 and $| \sX|$ are bounded as:
 \begin{align}
|B(x^\star,\delta)|&\geq \prod_{i=1}^m\left({\delta(n_i+1)\over 2ms_i} \right)^{s_i-1}
,\text{ and}\label{e:Bstar}\displaybreak[3]\\
| \sX|&\leq\prod_{i=1}^m(n_i+1)^{s_i-1}.
\end{align}

Now,
\begin{align*}
\E_\pi[\P(x)]	&\geq \P(x^\star) -\delta\lambda+ {1\over\beta}\log\left({|B(x^\star,\delta)|\over| \sX|}\right)\displaybreak[3]\\
		&\geq \P(x^\star) -\delta\lambda+ {1\over\beta}\log\left({\prod_{i=1}^m\left({\delta(n_i+1)\over 2ms_i} \right)^{s_i-1}   \over \prod_{i=1}^m(n_i+1)^{s_i-1}}\right)\displaybreak[3]\\
		&=\P(x^\star) -\delta\lambda+{1\over\beta} \log\prod_{i=1}^m \left({\delta\over 2ms_i}\right)^{s_i-1}\displaybreak[3]\\
		&\geq\P(x^\star) -\delta\lambda+{m(s-1)\over\beta} \log \left({\delta\over 2ms}\right)
\end{align*}
\normalsize
Consider two cases: (i) $\lambda\leq\eps/4,$ and (ii) $\lambda>\eps/4$.  For case (i), choose $\delta=1$ and let $\beta\geq {4m(s-1)\over\eps}\log 2 ms$.   Then,
\begin{align*}
\E_\pi[\P(x)] &\geq \P(x^\star) -\delta\lambda+{m(s-1)\over\beta} \log \left({\delta\over 2ms}\right)\displaybreak[3]\\
&\geq \P(x^\star) -\eps/4 - {m(s-1)\over\beta} \log 2ms\displaybreak[3]\\
&\geq \P(x^\star) -\eps/4 - {\eps m(s-1)\over 4m(s-1)\log 2ms}\log 2ms\displaybreak[3]\\
&=\P(x^\star) - \eps/2
\end{align*}
\normalsize

For case (ii),  note that $\lambda>\eps/4 \implies\delta=\eps/4\lambda<1$ so we may choose $\delta = \eps/4\lambda.$  Let $\beta\geq {4m(s-1)\over \eps}\log\left({8\lambda ms\over \eps} \right).$  Then
\small
\begin{align*}
\E_\pi[\P(x)] &\geq \P(x^\star) -\delta\lambda+{m(s-1)\over\beta} \log \left({\delta\over 2ms}\right)\displaybreak[3]\\
&=\P(x^\star) -\eps/4+{m(s-1)\over\beta} \log \left({\eps\over 8\lambda ms}\right)\displaybreak[3]\\
&=\P(x^\star) -\eps/4-{m(s-1)\over\beta} \log \left({8\lambda ms\over\eps}\right)\displaybreak[3]\\
&\geq\P(x^\star) -\eps/4-{\eps m(s-1)\over  4m(s-1) \log\left({8\lambda ms\over \eps} \right) } \log\left({8\lambda ms\over\eps}\right)\displaybreak[3]\\ 
&=\P(x^\star) -\eps/2
\end{align*}
\normalsize
as desired.\hfill\qed

\begin{lemma}\label{l:mixing time} 
For the Markov chain defined by modified log-linear learning with kernel $M$ and stationary distribution $\pi$, if the number of players within each population satisfies condition (ii) of Theorem~\ref{t:main theorem 1}, and $t$ is sufficiently large as in \eqref{e:time requirement}, then
\begin{equation}\label{e:distToS}\| \mu(t) - \pi\|_{TV}\leq \eps/2.\end{equation}
\end{lemma}

\noindent\emph{Proof:}
We begin by establishing a lower bound on the Sobolev constant for the Markov chain, $M$. We claim that, for the Markov chain $M$ defined in Appendix~\ref{a:M defn}, if $\P:  \sX\rightarrow [0,1]$ and
$m+\sum_{i=1}^m n_i^2 \geq \sigma$, then
\begin{equation}\label{e:rho bound}
\rho(M)\geq {e^{-3\beta}\over c_1m(m(s-1))!^2 n^2}
\end{equation}
for some constant $c_1$ which depends only on $s$.  Then,  from \eqref{e:mix time bounds}, a lower bound on the Sobolev constant yields an upper bound on the mixing time for the chain $M$.

Using the technique of \cite{Shah2010}, we compare the Sobolev constants for the chain $M$ and a similar random walk on a convex set. The primary difference is that our proof accounts for dependencies on the number of populations, $m$, whereas theirs considers only the $m=1$ case. As a result, our state space is necessarily larger. 
We accomplish this proof in four steps. In step 1, we define $M^\star$ to be the Markov chain $M$ with $\beta = 0,$ and establish the bound $\rho(M)\geq e^{-3\beta}\rho(M^\star).$ In step 2, we define a third Markov chain, $M^\dag$, and establish the bound $\rho(M^\star)\geq{1\over s}\rho(M^\dag).$ Then, in step 3, we establish a lower bound on the Sobolev constant of $M^\dag.$ Finally, in step 4, we combine the results of the first three steps to establish \eqref{e:rho bound}. 
We now prove each step in detail.

\noindent\tb{Step 1, $M$ to $M^\star$:}
Let $M^\star$  be the Markov chain $M$ with $\beta = 0,$ and let $\pi^\star$ be its stationary distribution. In $M^\star$ an updating agent chooses his next action uniformly at random. Per Equation \eqref{e:stationary distribution} with $\beta=0$, the stationary distribution $\pi^\star$ of $M^\star$ is the uniform distribution.
  Let $x,y\in  \sX.$  We bound $\pi_x/\pi_x^\star$ and $M(x,y)/M^\star(x,y)$ in order to use Corollary 3.15 in \cite{Montenegro2006}:
\begin{align*}
\frac{\pi_{x}}{\pi_{x}^\star} &= \frac{e^{\beta\P(x)}}{\sum_{y\in  \sX} e^{\beta\P(y)}}\cdot\frac{\sum_{y\in  \sX} e^0}{e^0}=\frac{| \sX|e^{\beta\P(x)}}{\sum_{y\in  \sX} e^{\beta\P(y)}}
\end{align*}
Since $\P(x)\in [0,1]$ for all $ x\in  \sX,$ this implies
\begin{equation}\label{e:ratio bound 1}
e^{-\beta}\leq\frac{\pi_{x}}{\pi_{x}^\star}\leq e^\beta
\end{equation}
Similarly, for $y = (\bx_{j}+\frac{1}{n}(e_j^k - e_j^{\ell}),\bx_{-j})$,
\begin{align*}
\frac{M(x,y)}{M^\star(x,y)} &= 
 \frac{s_je^{\beta\P(y)}}{\sum_{r = 1}^{s_j}e^{\beta\P(\bx_{j} + \frac{1}{n}(e^k _i- e^r_i),\bx_{-j})}}
\end{align*}
Since $\P(x)\in [0,1]$ for all $ x\in \sX,$
  for any $x,y\in \sX$ of the above form,
\begin{equation}\label{e:ratio bound 2}
e^{-\beta}\leq \frac{M(x,y)}{M^\star(x,y)}\leq e^\beta.
\end{equation}
For a transition to any $y$ not of the form above, $M(x,y) = M^\star(x,y) = 0.$  Using this fact and Equations \eqref{e:ratio bound 1} and \eqref{e:ratio bound 2}, we apply Corollary 3.15 in \cite{Montenegro2006}: 
\begin{equation}\label{Sob const ineq 1}\rho(M)\geq e^{-3\beta}\rho(M^\star).\end{equation}

\noindent\tb{Step 2, $M^\star$ to $M^\dag$:}
Consider the Markov chain $M^\dag$ on $ \sX $, where transitions from state $x$ to $y$ occur as follows:

\begin{noindlist}
\item Choose a population $N_j$ with probability $s_j/\sigma$
\item Choose  $k\in \{1,\ldots,s_j-1\}$ and choose $\kappa\in \{-1,1\}$, each uniformly at random.
\begin{itemize}
\item If $\kappa = -1$ and $\bx_{j}^k>0$, then $y = (\bx_{j} + \frac{1}{n}(e_{j}^{s_j} - e_j^k),\bx_{{-k}}).$ 
\item If $\kappa = 1$ and $\bx_{j}^{s_j}>0$, then $y = (\bx_{j} + \frac{1}{n}(e_j^k - e_{j}^{s_j}),\bx_{{-j}}).$ 
\end{itemize}
\end{noindlist}
Since $M^\dag(x,y) = M^\dag(y,x)$ for any $x,y\in  \sX$, $M^\dag$ is reversible with the uniform distribution over $ \sX$.  Hence the stationary distribution is uniform, and $\pi^\dag = \pi^\star$.

For a transition $x$ to $y$ in which an agent from population $N_j$ changes his action, $M^\star (x,y)\geq \frac{1}{s_j}M^\dag(x,y),$ implying 
\begin{equation}\label{e:ratio 3}
M^\star(x,y)\geq \frac{1}{s}M^\dag(x,y),\quad \forall x,y \in  \sX
\end{equation}
since $s\geq s_j,\;\forall i\in \{1,\ldots,m\}$.  Using \eqref{e:ratio 3} and the fact that $\pi^\star = \pi^\dag$, we apply Corollary 3.15 from \cite{Montenegro2006}:
\begin{equation}\label{Sob const ineq 2}\rho(M^\star)\geq \frac{1}{s}\rho(M^\dag)\end{equation}

\noindent\tb{Step 3, $M^\dag$ to a random walk:}\label{s:random walk}
The following random walk on 
\small
$$C = \left\{(z_1,\ldots,z_m)\in \Z_+^{\sigma-m}\st z_j\in \Z_+^{s_j-1},\;\sum_{k=1}^{s_j-1} z_{j}^k \leq n_j,\, \forall j\right\}$$
\normalsize
is equivalent to $M^\dag$.  
Transition from $x\to y$ in $C$ as follows:
\begin{noindlist}
\item Choose $j\in [\sigma-m]$ and $\kappa\in\{-1,1\}$, each uniformly at random
\item $y = \left\{\begin{array}{ll}x + \kappa e_j & \text{if } x+\kappa e_j \in C\\
							x&\text{otherwise}\end{array}\right.$.
\end{noindlist}

The stationary distribution of this random walk is uniform.  
We lower bound the Sobolev constant, $\rho(M^\dag),$ which, using the above steps, lower bounds $\rho(M)$ and hence upper bounds the mixing time of our algorithm.

Let $g:C\to \R$ be an arbitrary function.  To lower bound $\rho(M^\dag)$, we will lower bound $\mathcal{E}(g,g)$ and upper bound $\mathcal{L}(g)$.  The ratio of these two bounds in turn lower bounds the ratio $\mathcal{E}(g,g)/\mathcal{L}(g)$; since $g$ was chosen arbitrarily this also lower bounds the Sobolev constant. 
We will use a theorem due to \cite{Frieze1998} which applies to an extension of a function $g:C\to \R$ to a function defined over the convex hull of $C$; here we define this extension. 

Let $K$ be the convex hull of $C$.  Given $g: C\rightarrow \R$, we follow the procedure of \cite{Frieze1998, Shah2010} to extend $g$ to a function $g_\varepsilon: K\rightarrow \R$.  For $x\in C$, let $C(x)$ and $C(x,\eps)$ be the $\sigma-m$ dimensional cubes of center $x$ and sides 1 and $1-2\eps$ respectively.  For sufficiently small $\varepsilon>0$ and $z\in C(x)$, define $g_\eps : K\rightarrow \R$ by:
\begin{equation*}\label{e:feps defn}
g_\eps(z) := \left\{\begin{array}{ll}g(x) & \text{if } z\in C(x,\varepsilon)\\
\frac{(1+\eta(z))g(x) + (1-\eta(z))g(y)}{2} & \text{otherwise } \end{array}\right.
\end{equation*}
\normalsize
where $y\in C$ is a point such that $D := C(x)\cap C(y)$ is the closest face of $C(x)$ to $z$ (if more than one $y$ satisfy this condition, one such point may be chosen arbitrarily), and $\eta := \frac{\dist(z,D)}{\varepsilon}\in[0,1).$  The $\dist$ function represents standard Euclidean distance in $\R^{\sigma-m}.$   

Define
\small
\begin{align}
I_{\eps} &:= \int_K \bigl|\nabla g_\eps(z)\bigr|^2dz\\
J_{\eps} &:= \int_K g_\eps(z)^2\log\frac{g_\eps(z)^2\vol(K)}{\int_K g_\eps(y)^2dy}dz. \label{e:Jeps}
\end{align}
\normalsize
Applying Theorem 2 of \cite{Frieze1998} for $K\in \R^{\sigma-m}$ with diameter $\sqrt{\sum_{i=1}^m n_i^2}$, if $m+\sum_{i=1}^mn_i^2\geq\sigma,$
\begin{equation}\label{IJ inequality}\frac{\eps I_\eps}{J_\eps}\geq \frac{1}{A\sum_{i=1}^m n_i^2}.\end{equation}
We lower bound $\mathcal{E}(g,g)$ in terms of $\eps I_\eps$ and then upper bound $\mathcal{L}(g)$ in terms of $J_\eps$ to obtain a lower bound on their ratio with Equation \eqref{IJ inequality}.  
The desired lower bound on the Sobolev constant follows. 

Using similar techniques to \cite{Shah2010}, we lower bound $\mathcal{E}(g,g)$ in terms of $\eps I_\eps$ as
\small
\begin{align}
I_\eps 	
		 &\leq\frac{|C|(\sigma - m)}{\eps}\mathcal{E}(g,g) + O(1)\displaybreak[3].\nonumber
\end{align}
\normalsize
Then,
$\eps I_\eps \;\mathop{\leq}_{\eps\rightarrow 0} \;|C|(\sigma - m)\mathcal{E}(g,g)$, and hence 
\begin{equation}\label{I inequality}\mathcal{E}(g,g)\;\mathop{\geq}_{\eps\rightarrow 0}\;\frac{\eps I_\eps}{|C|(\sigma-m)}.\end{equation}
Again, using similar techniques as \cite{Shah2010}, we bound $J_\eps$ as
\small
\begin{align}
\frac{J_\eps}{\vol(K)}	
					&\geq\frac{|C|}{2^{2(\sigma-m)}\vol(K)(\sigma-m)!^2}\mathcal{L}(f).\nonumber
\end{align}
Then 
\small
\begin{equation}\label{J inequality}\mathcal{L}(g)\mathop{\leq}_{\eps\to 0} \frac{2^{2(\sigma-m)}(\sigma-m)!^2}{|C|}J_\eps.\end{equation}
\normalsize

\noindent\tb{Step 4, Combining inequalities:}
Using inequalities \eqref{IJ inequality}, \eqref{I inequality}, and \eqref{J inequality}, 
\small
\begin{equation}\frac{\mathcal{E}(f,f)}{\mathcal{L}(f)}\geq \frac{1}{2^{2(\sigma-m)}A(\sigma-m)(\sigma - m)!^2\sum_{i = 1}^m n_i^2},\end{equation}
\normalsize
$\forall f: C\rightarrow\R.$  Therefore,
\small
\begin{align}
\rho(M^\dag) &= \min_{f: C\rightarrow \R}\frac{\mathcal{E}(f,f)}{\mathcal{L}(f)}\displaybreak[3]\nonumber\\
&\geq \frac{1}{2^{2(\sigma-m)}A(\sigma-m)(\sigma - m)!^2\sum_{i = 1}^m n_i^2}\label{Sob const ineq 3}
\end{align}
\normalsize

Combining equations \eqref{Sob const ineq 1}, \eqref{Sob const ineq 2}, and \eqref{Sob const ineq 3}
\small
$$\rho(M)\geq {e^{-3\beta}\over 2^{2ms} c_1m^2(m(s-1))!^2 n^2}$$
\normalsize
 where $c_1$ is a constant depending only on $s$.  
 
From here, Lemma~\ref{l:mixing time} follows by applying Equation \eqref{e:mix time bounds} in a similar manner as the proof of Equation (23) in \cite{Shah2010}. The main difference is that the size of the state space is bounded as $|\sX|\leq\prod_{i=1}^m(n_i+1)^{s_i+1}$ due to the potential for multiple populations.
 \hfill\qed

Combining Lemmas~\ref{l:stationary distribution} and \ref{l:mixing time} results in a bound on the time it takes for the expected potential to be within $\eps$ of the maximum, provided $\beta$ is sufficiently large.
The lemmas and method of proof for Theorem \ref{t:main theorem 1} follow the structure of the supporting lemmas and proof for Theorem 3 in \cite{Shah2010}. The main differences have arisen due to the facts that i) our analysis considers the multi-population case, so the size of our state space cannot be reduced as significantly as in the single population case of \cite{Shah2010}, and ii) update rates in our algorithm depend on behavior within each agent's own population, instead of on global behavior.

\smallskip

\noindent\emph{Proof of Theorem~\ref{t:main theorem 1}}:\\
From Lemma \ref{l:stationary distribution},
if condition (i) of Theorem~\ref{t:main theorem 1} is satisfied and $\beta$ is sufficiently large as in \eqref{e:beta lb},
then 
$\E_\pi[\P(x)]\geq\max_{x\in \sX}\P(x)-\eps/2.$ 
From Lemma \ref{l:mixing time}, if condition (ii) of Theorem~\ref{t:main theorem 1} is satisfied, and $t$ is sufficiently large as in \eqref{e:time requirement},
then 
$\|\mu(t) - \pi\|_{TV}\leq \eps/2.$
Then
\begin{align*}
\E[\P(a(t)|_{\sX})] 	&=\E_{\mu(t)}[\P(x)] \nonumber\displaybreak[3]\\
			&\geq \E_{\pi}[\P(x)] - \|\mu(t) - \pi\|_{TV}\cdot \max_{x\in  \sX}\P(x)\nonumber\displaybreak[3]\\
&\stackrel{(a)}{\geq} \max_{x\in  \sX}\P(x) - \eps \nonumber\displaybreak[3]
\end{align*}\
where (a) follows from \eqref{e:expPot}, \eqref{e:distToS}, and the fact that $\P(x)\in[0,1].$
\qed

\begin{IEEEbiography}[{\includegraphics[width=1in,height=1.25in,clip,keepaspectratio]{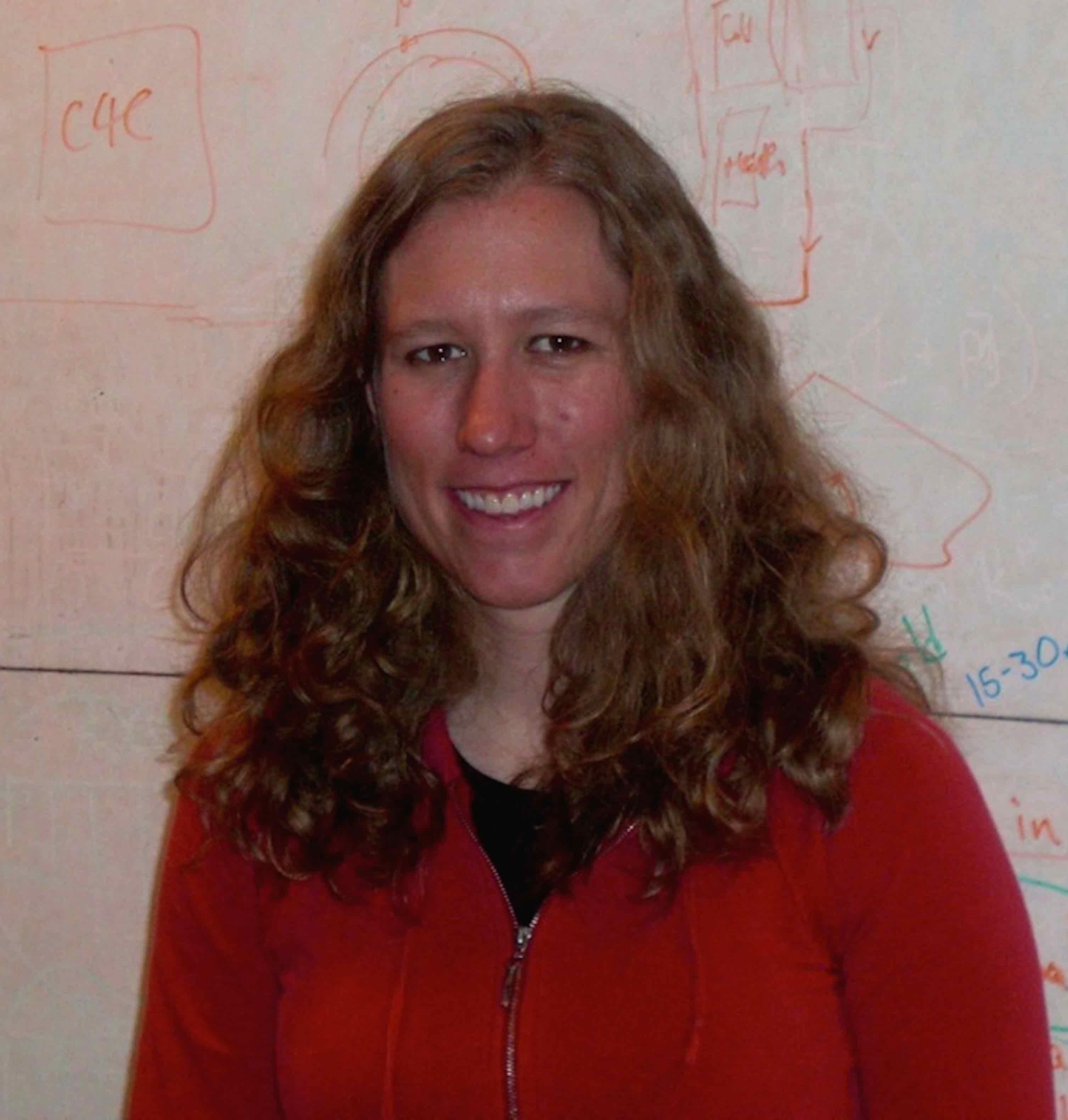}}]{Holly Borowski} is a Ph.D. candidate in the Aerospace Engineering Sciences Department at the University of Colorado, Boulder. Holly received a BS in Mathematics in 2004 at the US Air Force Academy. She received the NASA Aeronautics graduate scholarship (2012), the Zonta International Amelia Earhart scholarship (2014), and the Philanthropic Educational Organization scholarship (2015). Holly's research focuses on the role of information in the control of distributed multi-agent systems.
\end{IEEEbiography}

\begin{IEEEbiography}[{\includegraphics[width=1in,height=1.25in,clip,keepaspectratio]{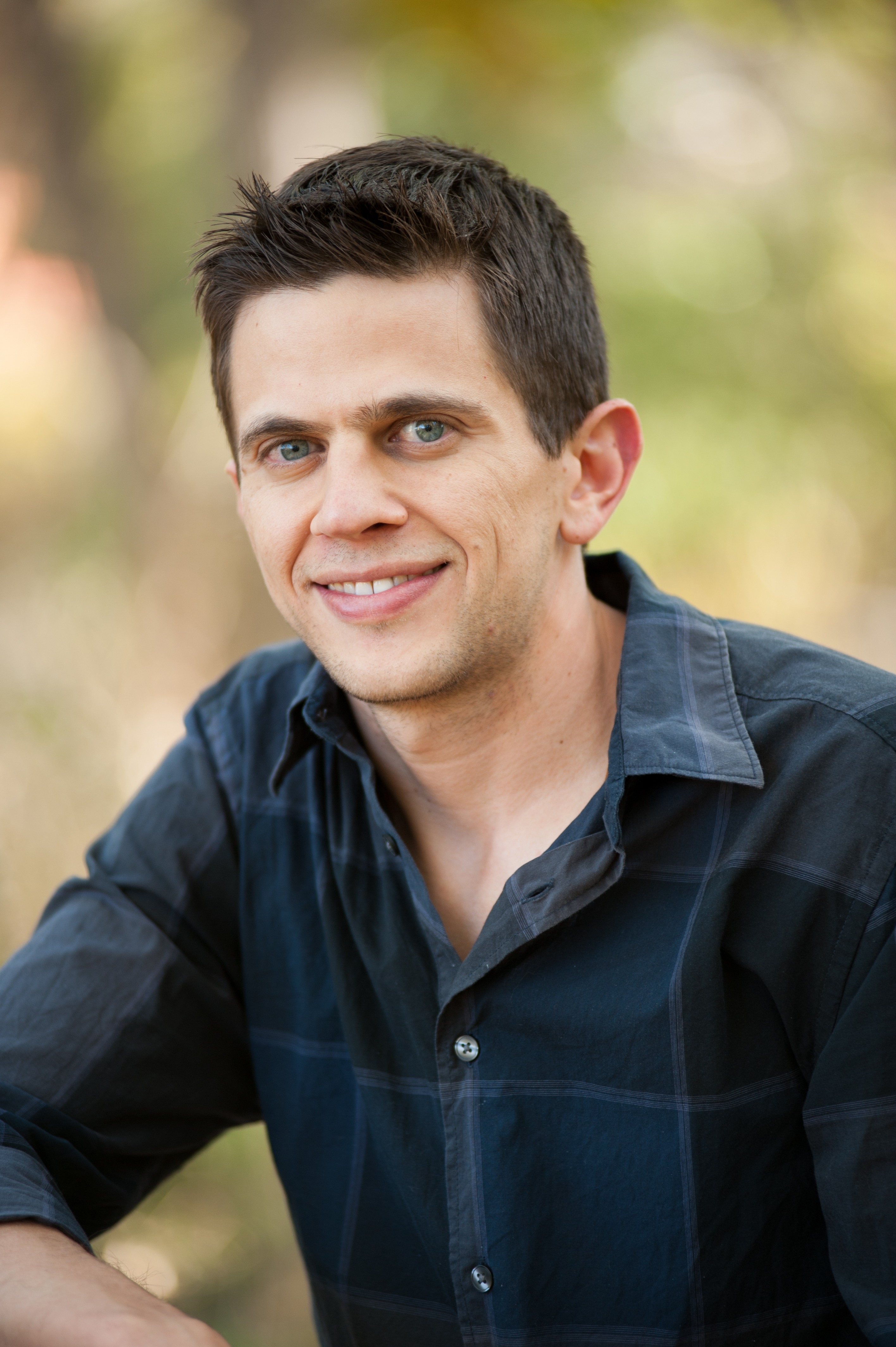}}]{Jason Marden} is an Assistant Professor in the Department of Electrical and Computer Engineering at the University of California, Santa Barbara.  Jason received a BS in Mechanical Engineering in 2001 from UCLA, and a PhD in Mechanical Engineering in 2007, also from UCLA, under the supervision of Jeff S. Shamma, where he was awarded the Outstanding Graduating PhD Student in Mechanical Engineering.  After graduating from UCLA, he served as a junior fellow in the Social and Information Sciences Laboratory at the California Institute of Technology until 2010 when he joined the University of Colorado.  Jason is a recipient of the NSF Career Award (2014), the ONR Young Investigator Award (2015), the AFOSR Young Investigator Award (2012), and the American Automatic Control Council Donald P. Eckman Award (2012).   Jason's research interests focus on game theoretic methods for the control of distributed multi-agent systems.
\end{IEEEbiography}


\end{document}